\documentclass[fleqn,10pt]{wlscirep}
\usepackage[utf8]{inputenc}
\usepackage[T1]{fontenc}
\title{Mining urban lifestyles: urban computing, human behavior and recommender systems}

\author[1]{Sharon Xu}
\author[2,3]{Riccardo Di Clemente}
\author[4,5*]{Marta C. Gonz\'alez}
\affil[1]{Department of Civil and Environmental Engineering, Massachusetts Institute of Technology, Cambridge, MA 02139, USA}
\affil[2]{Centre for Advanced Spatial Analysis, University College London, London WC1E 6BT, GBR}
\affil[3]{Department of Computer Science, University of Exeter, Exeter EX4 4PY, GBR}
\affil[4]{Department of City and Regional Planning, UC Berkeley, 406 Wurster Hall, Berkeley, 94720, USA}
\affil[3]{Lawrence Berkeley National Laboratory, Cyclotron Road, Berkeley, 94720, USA}

\affil[*]{martag@berkeley.edu}

\begin{document}

\flushbottom
\maketitle
%
%
\noindent{\parbox{\dimexpr\linewidth-2\fboxsep\relax}{\color{color1}\normalsize\sffamily{This  is a postprint of the 5th chapter published in the book Big Data Recommender Systems - Volume 2: Application Paradigms  and is subject to Institution of Engineering and Technology Copyright. The copy of record is available at the IET Digital Library  \href{http://doi.org/10.1049/PBPC035G_ch5}{DOI: 10.1049/PBPC035G{\_}ch5}}}}
\vskip10pt

In the last decade, the digital age has sharply redefined the way we study human behavior. With the advancement of data storage and  sensing technologies, electronic records now encompass a diverse spectrum of human activity, ranging from location data \cite{cmobility_limits,cmobility}, phone \cite{ccall1,ccall2} and email communication \cite{poissonianemail} to Twitter activity \cite{ctwitter} and open-source contributions on Wikipedia and OpenStreetMap \cite{cwikipedia,copenstreetmap}. In particular, the study of the shopping and mobility patterns of individual consumers has the potential to give deeper insight into the lifestyles and infrastructure of the region. 
Credit card records (CCRs) provide detailed insight into purchase behavior and have been found to have inherent regularity in consumer shopping patterns \cite{krumme2013predictability}; call detail records (CDRs) present new opportunities to understand human mobility \cite{marta}, analyze wealth \cite{wealth}, and model social network dynamics \cite{morse}.\\

Regarding the analysis of CDR data, there exists a wide body of work characterizing human mobility patterns. As a notable example, \cite{marta} describes the temporal and spatial regularity of human trajectories, showing that each individual can be described by a time independent travel distance and a high probability of returning to a small number of locations. Further, the authors are able to model individual travel patterns using a single spatial probability distribution. There has also been work at the intersection of similar datasets, such as the inference of friendships from mobile phone data \cite{pentland1}, or the analysis such data in relation to metrics on spending behavior such as diversity, engagement, and loyalty \cite{pentland2}. Recent work \cite{rdicle} uses the Jaccard distance as a similarity measure on motifs among spending categories, then applies community detection algorithms to find clusters of users. These studies propose models for either mobility or spending behavior, but not in conjunction.\\

The only known paper that incorporates both aspects \cite{mining} frames its analysis only on an aggregate scale of city regions. However, the coupled collaborative filtering methods (also known as collective matrix factorization) used in \cite{mining} have been successfully applied in a variety of urban computing applications for data fusion and prediction \cite{zhengoverview,zhengurbcompoverview,zhengtrajectoryoverview}, from location-based activity recommendations \cite{zhengactivityrec,zhengactivityrec2} to travel speed estimation on road segments \cite{cmfpollution}. Recent work includes methods that use Laplacian regularization \cite{laplacian} to leverage social network information, and use geometric deep learning matrix completion methods to model nonlinearities \cite{deeplearning}.\\

In this chapter, we jointly model the lifestyles of individuals, a more challenging problem with higher variability when compared to the aggregated behavior of city regions. Using collective matrix factorization, we propose a unified dual view of lifestyles. Understanding these lifestyles will not only inform commercial opportunities, but also help policymakers and nonprofit organizations understand the characteristics and needs of the entire region, as well as of the individuals within that region. The applications of this range from targeted advertisements and promotions to the diffusion of digital financial services among low-income groups.

\subsection*{Mining Shopping and Mobility Patterns}
Location and transactional data offer valuable perspectives on the lifestyles of each user. For example, we may expect the shopping purchases of middle-aged parents to include groceries and fuel, while their mobility patterns may center around localities near home and work locations, in addition to points of interest such as supermarket, laundry, and so on. We use mobility information to aid in the prediction of shopping behavior, connecting the two views using collective matrix factorization \cite{gordon}. In this way, we discover representative patterns relating shopping and mobility, characterizing behavior for a richer understanding into urban lifestyles and improved prediction of behavior.\\

The high granularity of such digital records allows modeling at the level of the individual, providing a new framework in which to relate movement and spending. However, in using CDR data for data on individuals, we must deal with issues of sparsity and lack of contextual information on the user's activities. In proposing this dual view of lifestyles, our contributions can be summarised as follows:\\[2ex]
\textit{Prediction of Shopping Behavior with Data Sparsity}\\
There are many individuals for which we have no CDR data. To deal with this data sparsity issue, we construct a framework that uses mobility patterns as supplementary information in the prediction of shopping behavior. We connect the two perspectives on lifestyles using collective matrix factorization (collective matrix factorization). In comparison to modeling only shopping behavior, we find that incorporating mobility information in the prediction of shopping lifestyles leads to a significant reduction in root mean square error (RMSE).\\[2ex]
\textit{Adding Contextual Information to Location Data}\\
We transform mobility data using external data sources to better relate CCR to CDR Data. Although CCRs provide high granularity at the level of the individual user, spatial granularity can range from a radius of 200 - 1000 meters, and there is no contextual information for the user's activities within that region. Thus, there has been little previous work leveraging CCR data for prediction with CDR data.\\[2ex]
\textit{Multi-Perspective Lifestyles}\\
We describe the mappings between shopping and mobility patterns, connecting the two views to provide a novel understanding of consumer behavior in urban regions.
\begin{figure}[h!]
\caption{Our framework}
\includegraphics[width=\textwidth]{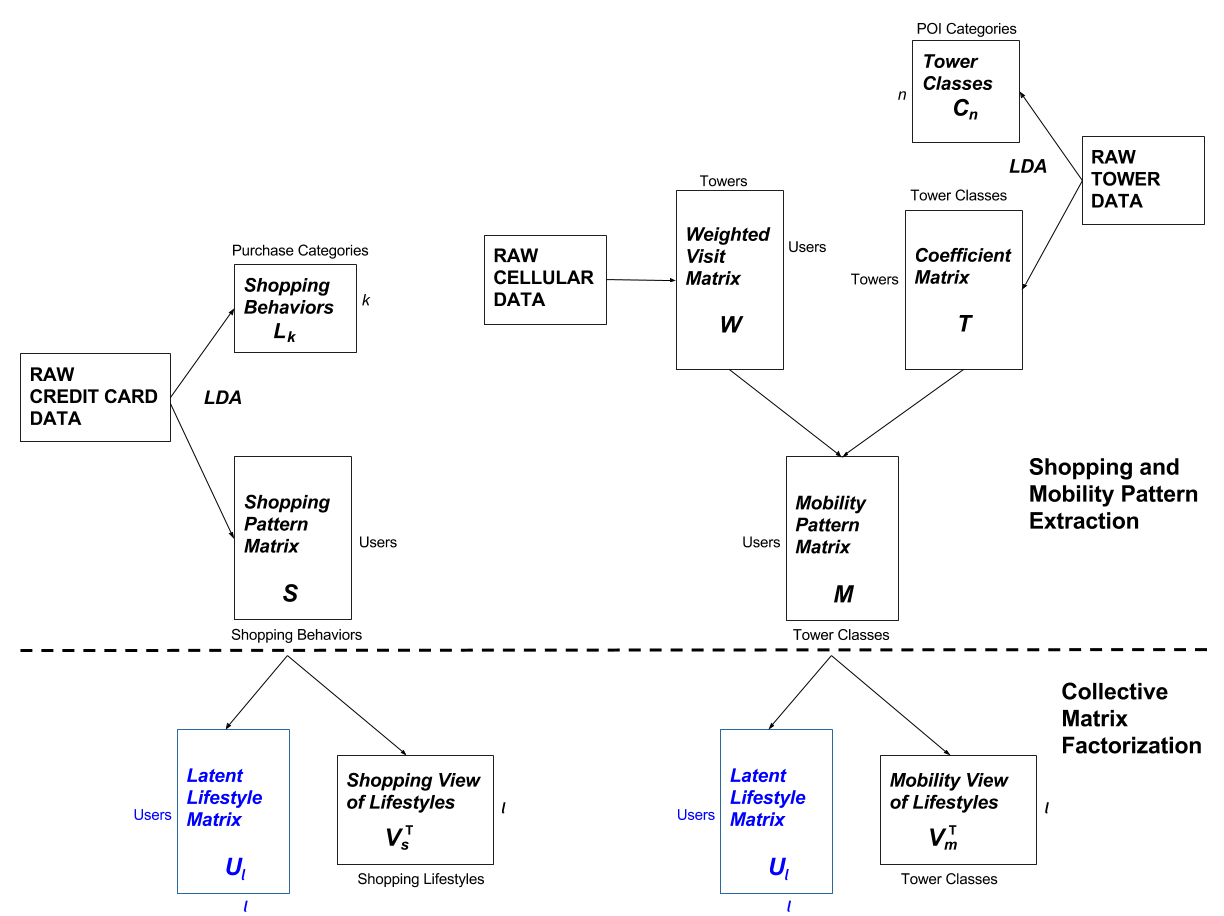}
\label{overview}
\end{figure}

\subsection*{Data}
The primary datasets used in this chapter consist of two sets of anonymised data for residents in Mexico throughout five months in 2015:
\begin{itemize}
\item Call detail records (CDRs). CDRs are produced with each telephone exchange, These kocation records give the nearest cellular tower at the time of a placed call. There are 1192 cell towers throughout Mexico City -- as users tend to visit a small subset of these towers, this mobility data is extremely sparse. In a count matrix denoting user visits to towers, 98\% of entries indicate zero visits.
\item Credit card records (CCRS). CCRs are recorded with each purchase and denote the purchase category, or  Merchant Category Code (MCC), of the transaction as well as the amount spent. Each month, we have on the order of 10 million financial transactions and 200 million location records.
\end{itemize}

\section*{Discovering Shopping Patterns}
Our spending habits reflect our lifestyles, capturing an essential aspect of our behavior. Within the computational social science community, the question remains whether pervasive trends exist among disparate groups at urban scale \cite{rdicle}. In this chapter, we use latent Dirichlet allocation (LDA) \cite{blei} to identify topics (behavioral patterns) among individuals, representing each individual's spending lifestyle as a finite mixture of an underlying set of behaviors. Each behavioral pattern, in turn, is modeled as a mixture of a set of words (Merchant Category Codes, or MCCs). These topics are determined by co-occurrences of words within a document. For example, in an article database, we may uncover a topic containing the words ``data'', ``processing'', ``computer'', and so on because these words frequently appear in an article together. \\[2ex]
By putting a Dirichlet prior on the per-user behavior distribution and per-behavior MCC distribution, LDA controls the sparsity of the number of topics per document (the number of behaviors per individual), as well as the number of words per topic (the number of MCCs per behavioral pattern). In this way, each individual is represented by a small number of behaviors, and each behavior involves making a small set of purchase categories with high frequency.\\[2ex]
As a generative model, LDA allows us to calculate the probabilities (assignments to shopping behaviors) of previously unseen users. We train the model on 40\% of the users, and generate the matrix $S$ for the remaining 60\%. In so doing, we set up the prediction of lifestyles for unseen users, assessing the LDA model itself in addition to the relation of shopping with mobility patterns. We experiment with the choice of number of behaviors to learn, as well as adding a categorical variable describing amount spent to each MCC. To maximise interpretability, we choose five topics while using MCCs as input only.\\[2ex]
In Fig. 2 we plot the twenty most highly weighted MCCs of the five shopping behaviors. The first shopping behavior describes credit card usage that is centered on food-related purchases such as Grocery Stores, Misc. Food Stores and Restaurants. The second shopping behavior seems to be associated primarily with business purchases, with spending within MCCs such as Fax Services and Financial Institutions. The third shopping behavior is dominated by relative ``luxuries'' such as purchases in the Cable and Department Store categories, and is characterised by a relatively high proportion of Air Travel and Hotel Lodging MCCs. The fourth shopping behavior contains primarily purchases in Computer Network Services and Service Stations (gas stations). The third and fourth shopping behavior describe a slightly wealthier portion of the population, as only 35\% of Mexicans owned a computer in 2010 \cite{computer}, and only 44.2\% own a car \cite{car}. Lastly, the fifth shopping behavior captures purchase primarily for toll fees and subscription services.

\begin{figure}[h!]
\centering 
\includegraphics[width=\textwidth]{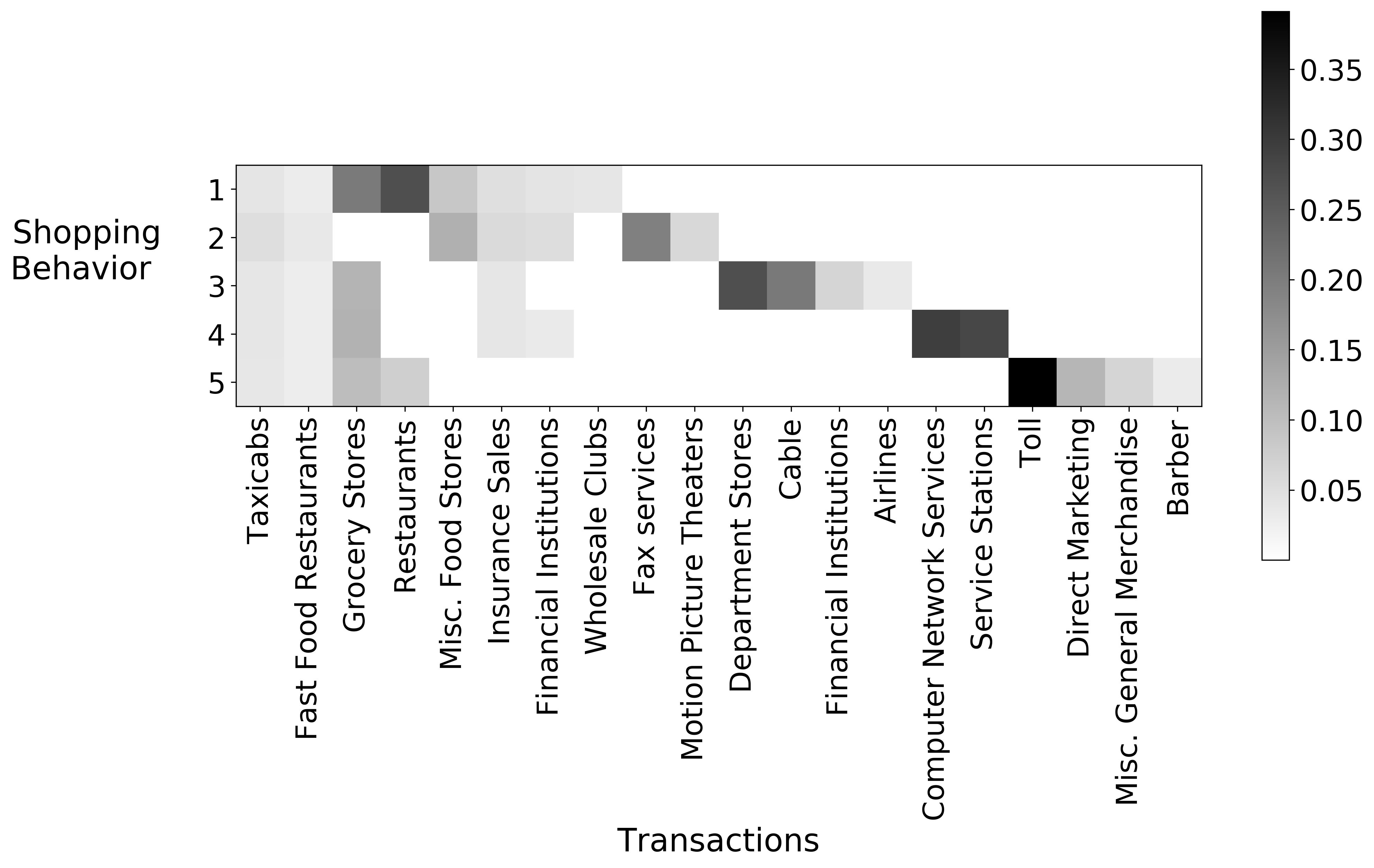}
\label{shopping}
\caption{The top weighted purchase categories of the five shopping behaviors learned from LDA.}
\end{figure}




\section*{Mobility Pattern Extraction}
\subsection*{Extracting Cellular Tower Location Types}
Within the CDR data, each tower is the site for a corresponding cell within the Voronoi diagram; i.e., it is the closest tower to any point within this cell. We define a ``visit'' to a cellular tower as a call placed within its corresponding cell. In order to relate cellular towers to spending behavior, for each tower we crawl Google's API for points of interest within a certain radius. To determine this radius, we use Delaunay triangulation, a widely used method in computational geometry. Delaunay triangulation gives the dual graph to the Voronoi diagram, maximizing the minimum angle among all the triangles within the triangulation, and connecting the sites in a nearest-neighbor fashion \cite{voronoi}. For each tower, we set the crawling radius to be half the average distance from the site to its neighbors.\\[2ex]
Treating each of the Voronoi cells as a document and the POI categories as words, we use latent Dirichlet allocation to discover underlying tower ``classes'' that will be more informative of shopping behavior. We remove from the vocabulary any POI categories that occur with over 25\% frequency. These removed categories are uninformative classifications such as ``point of interest'' and ``establishment''. For purposes of interpretability, we learn the LDA model with twenty classes on the 1192 towers. \\[2ex]
\begin{figure}[h!]
\caption{The top weighted POI categories of a subset of tower classes learned from LDA.}
\center \includegraphics[width=\textwidth]{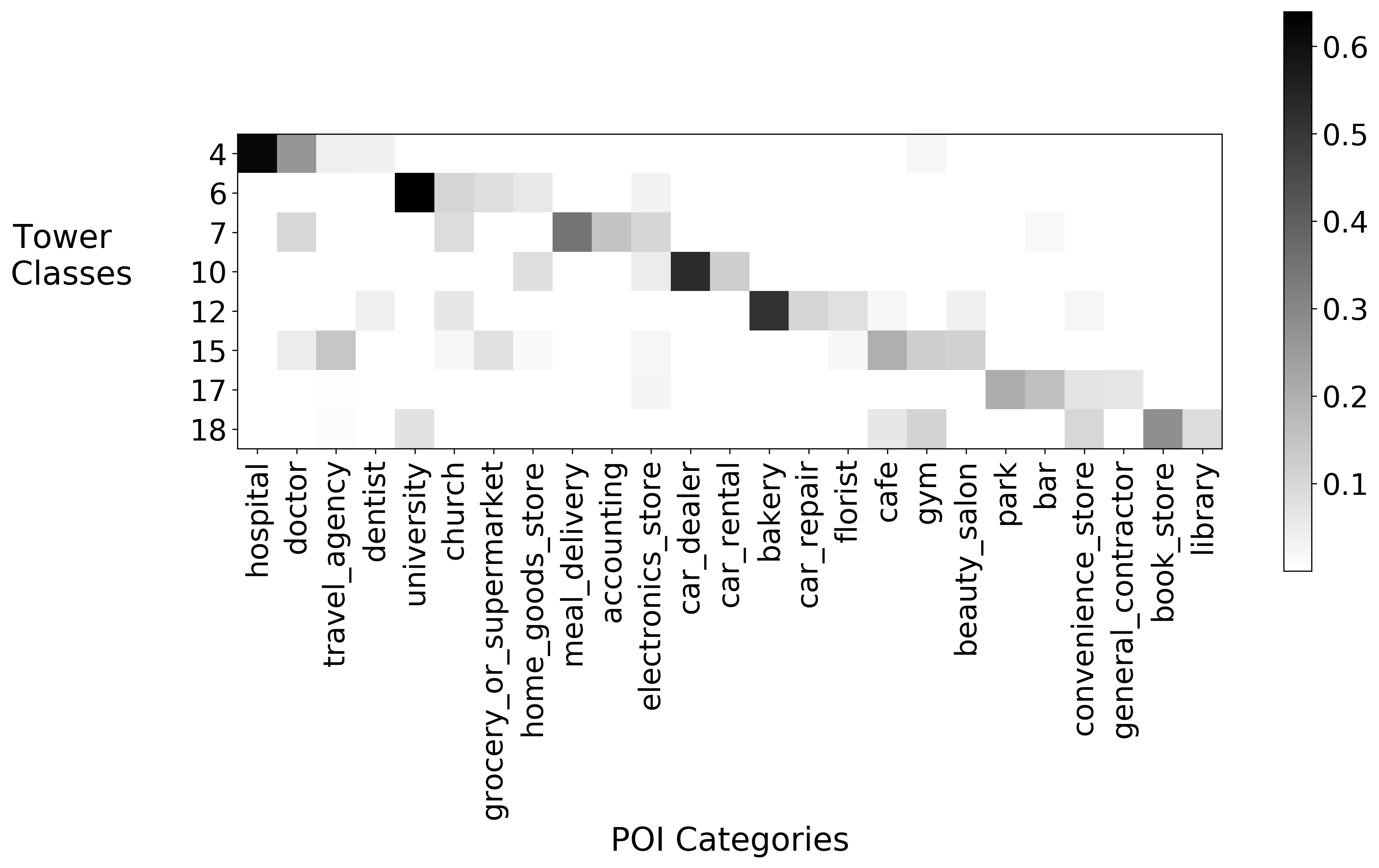}
\label{poi}
\end{figure}
In Fig. \ref{poi}, we show a subset of tower classes highly weighted within our final lifestyles (see Section 6), and the corresponding points of interest with the highest probabilities. From the sample topics in Fig. \ref{poitopics}, we see each tower class puts specific emphasis on related points of interest, such as ``hospital'' and ``doctor'', ``car rental'' and ``car repair'', or ``book store'' and ``library''. In this way, we cluster the towers in terms of nearby POI categories, obtaining contextual information more directly related to shopping.
\vspace{3em}
\begin{figure}[h!]
\centering
\caption{Sample topics from learned from LDA, treating each tower as a document and each POI as a word.}
\includegraphics[width=0.6\textwidth]{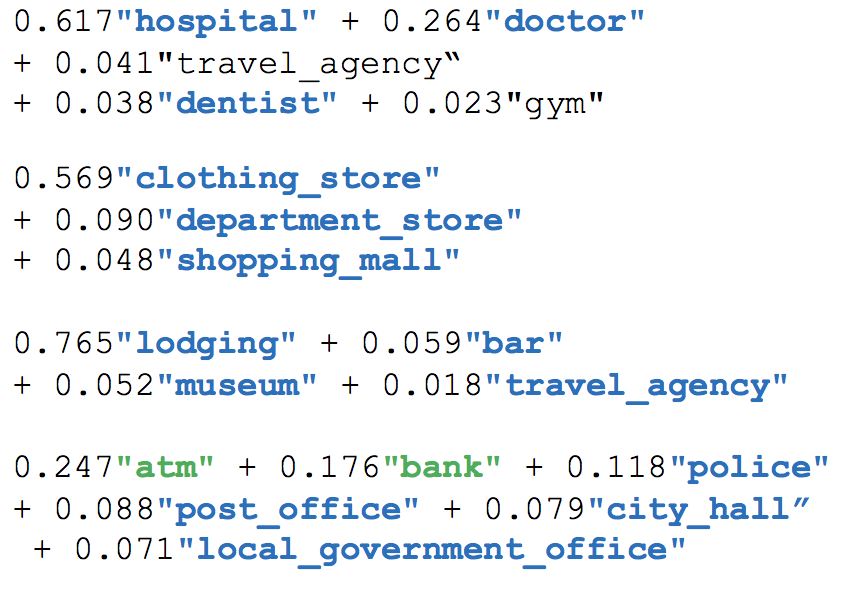}
\label{poitopics}
\end{figure}
\subsection*{Baseline Methods}
Before introducing our model, we present the results of several baseline methods, illustrating the challenges of incorporating CDR data into the prediction of shopping patterns.\\[2ex]
\textit{Regression on Average Amount Spent}\\
Using the columns of the per tower count matrix $W$ directly as features, we use regression with L1 regularization to predict the average amount spent by the user per week. As we increase regularization we increase the test R-squared, but due to a combination of sparsity and lack of signal achieve a maximum test R-squared of 0 as the coefficients shrink to 0. \\[2ex]
\textit{Classification of Primary Shopping Behavior}\\
For each user, we take as our outcome the highest weighted shopping behavior from the topic proportions learned from LDA. This is the user's primary behavior. Again using the columns of $W$ as our features, we employ a range of classifiers including SVM and AdaBoost to predict primary behavior. We find that the best classifier achieves only 21.6\% accuracy, when already 21.9\% of users fall into a single class.
\subsection*{Characterizing Mobility Patterns}
From the Voronoi diagram of the $p$ cell tower locations, we construct a matrix $W \in \mathbb{R}^{nxp}$ where each entry $w_{ij}$ is the number of days individual $i$ visited tower $j$ throughout five months. We weight these counts using TF-IDF, a common method for text representation \cite{tfidf}. Using TF-IDF, we offset the tower counts by the frequency of the tower in the data, so that a user's visit to an uncommonly visited tower is assigned a higher weight.
We now have a matrix $W \in \mathbb{R}^{nxp} $ that characterises users in terms of tower visits, and a matrix $C_m \in \mathbb{R}^{pxd}$, where $d$ is the chosen number of tower classes. We define our mobility pattern matrix as $ M = WT$, achieving a significant dimensionality reduction with $M \in \mathbb{R}^{nxd}$. In this manner, we obtain a representation of mobility more closely related to shopping behavior, as users are now characterised by their visits to tower classes defined by POI categories.
\section*{Predicting Shopping Behavior}
For many users, we have access to data on mobility patterns ($M$) but not shopping patterns ($S$). In this section, we describe our methodology for incorporating mobility information in addition to shopping information for the matrix completion problem of predicting the shopping behavior of unseen users.
\subsection*{Collective Matrix Factorization}
We denote $S$ as the matrix of behavior proportions obtained from latent Dirichlet allocation, and $M$ as the matrix of weighted visit frequencies to the different tower classes. Modeling each user's shopping and mobility behavior as two views of the same lifestyle, we assume that $S$ and $M$ are generated from a matrix $U_l$ containing the latent lifestyle information of each user. 
\begin{align*}
S &\approx U_l V_s^T\\[1.5ex]
M &\approx U_l V_m^T
\end{align*}
Traditionally, the objective function under this model is represented as 
$$ \mathcal{L}(U_l, V_s, V_m) \; = \; ||S - U_l V_s^T||^2 + ||M - U_l V_m^T||^2 + \lambda _1||U_l||^2 + \lambda _2||V_s||^2 + \lambda _3||V_m||^2 $$
In this chapter, we use group-wise sparse collective matrix factorization \cite{gcmf}, which puts group-sparse priors following $ \mathcal{N}(0, \sigma _k^2) $ on the columns of matrices $V_s$ and $V_m$, where the columns are the groups indexed by $k$ and $\sigma _k^2 $ is small. This allows the matrix to learn private factors for the relation between latent lifestyles ($U_l$) and the shopping aspect ($V_s$), and correspondingly between latent lifestyles ($U_l$) and the mobility aspect ($V_m$). More specifically, if the $k^{th}$ column of $V_m$ is null, then the $k^{th}$  factor impacts only the shopping pattern matrix $S$.
\section*{Results}
\subsection*{Prediction}
In our problem, credit card data is unknown for many users, but we would like to use mobility information to predict their shopping behavior; i.e., $S$ contains many empty rows. Thus, to test the performance within this setting, we remove rows from the shopping behavior $S$ to predict the shopping behavior of users for which we have no credit card information. We use 10-fold cross validation and compare our collective matrix factorization predictions with the actual values. 
We use the popular metric root mean square error (RMSE) to evaluate our model.
$$\mathrm{RMSE} = \sqrt{ \dfrac{1}{T} \sum\limits_{i,j} (S_{i,j} - \hat{S_{i,j}})^2 } $$

Using cross-validation to determine the rank (number of lifestyles), we find that the inclusion of mobility data leads to a 1.3\% decrease in RMSE and obtain a test error of 21.6\%.

\subsection*{Dual Lifestyles}

Using collective matrix factorization, we also obtain both the dual shopping and mobility views of these latent lifestyles, in $V_s$ and $V_m$ respectively. \\[2ex]
\textbf{Lifestyle 1} is connected with wealthier shopping behavior typical common to urban white collars. The top weighted shopping patterns indicate spending on cable, air travel, hotels and at department stores as well as gas stations and computer network services (Fig. 2: behaviors 3 and 4, respectively). This suggests that people who can afford to spend on relative luxuries tend to have vehicles and thus higher mobility, visiting a wider range of tower classes. The mobility patterns of this lifestyle focus on areas with points of interest such as universities, accounting, electronics, bakeries and car repair (Fig. \ref{poi}: tower classes 6, 7, 12, 17 and 20).\\[2ex]
\textbf{Lifestyle 2} is extremely food-oriented, with high weight on shopping behavior 1. Mobility patterns suggest visits to cafes, gyms and convenience stores.\\[2ex]
\textbf{Lifestyle 3} primarily captures the transportation aspect of lifestyles. Top weighted mobility patterns indicate visits to areas with car rental and car repair (tower classes 10 and 12), while shopping patterns include gas stations in behavior 4 and food in behavior 1.\\

\section*{Discussion}

In this study, we relate the shopping and mobility patterns of consumers on an individual level for the first time. Viewing these as aspects of the same underlying lifestyle, we set up a framework to incorporate CDR data in the prediction of shopping patterns for unseen users. We achieve a significant increase in prediction and recover interesting relationships between shopping and mobility.\\

There are many directions for future work. In terms of modeling formulation, it would be interesting to introduce a temporal dimension into the task of shopping prediction, as human behavior and needs vary over time. There is also the opportunity to include social regularization in the collective matrix factorization formulation, constraining each user to be similar to his or her neighborhood. In addition, stronger prediction methods may be achieved by modeling nonlinear relationships using geometric deep learning methods described by \cite{deeplearning}.

\section*{Acknowledgements}
We thank Grandata for supplying the data. (For contractual and privacy reasons, the raw data cannot be provided.)

\bibliographystyle{vancouver-modified}
\bibliography{main_new} 

\end{document}